\documentstyle[11pt,newpasp,twoside,epsf,amssymb,amstext]{article}
\newcommand{\HIbold}{{H\footnotesize{\bf{I}}}}
\newcommand{\HI}{\mbox{\sc H{i}}}
\newcommand{\nhi}{\mbox{$N_{\rm HI}$}}
\newcommand{\Lya}{\mbox{Ly$\alpha$}}
\newcommand{\msun}{\mbox{$M_\odot$}}
\newcommand{\mhi}{\mbox{$M_{\rm HI}$}}
\newcommand{\kms}{\mbox{km s$^{-1}$}}
\newcommand{\cm}{\text{cm}$^{-2}$}

\def\edcomment#1{\iffalse\marginpar{\raggedright\sl#1\/}\else\relax\fi}
\reversemarginpar

\begin{document}
\title{Local Column Density Distribution Function from \HIbold\ selected galaxies.}
\author{Emma Ryan-Weber}
\affil{University of Melbourne, VIC 3010 Australia}
\author{Rachel Webster}
\affil{University of Melbourne, VIC 3010 Australia}
\author{Lister Staveley-Smith}
\affil{Australia Telescope National Facility, CSIRO, PO Box 76, 
Epping, NSW 1710, Australia}

\begin{abstract}
The cross-section of sky occupied by a particular neutral hydrogen
column density (\nhi) provides insight into the nature of \Lya\
absorption systems. We have measured this column density distribution
at z=0 using 21-cm \HI\ emission from a blind survey. A subsample of
HI Parkes All Sky Survey (HIPASS) galaxies have been imaged with the
Australia Telescope Compact Array (ATCA). The contribution of low \HI\
mass galaxies ($10^{7.5}$ to $10^8$ \msun) is compared to that of
M$_*$ ($10^{10}$ to $10^{10.5}$ \msun) galaxies. We find that the
column density distribution function is dominated by low \HI\ mass
galaxies in the range $3\times10^{18} < \nhi\ < 2\times10^{20} $
cm$^{-2}$. This result is not intuitively obvious. M$_*$ galaxies may
contain the bulk of the \HI\ gas, but the cross-section presented by
low \HI\ mass galaxies ($10^{7.5}$ to $10^8$ \msun) is greater at
moderate column densities. This result implies that moderate \nhi\
\Lya\ absorption systems may be caused by a range of galaxy types
and not just large spiral galaxies as originally thought.
\end{abstract}

\section{Introduction}
Lyman $\alpha$ (\Lya) absorption systems provide a powerful probe into the
distribution of baryonic matter in the Universe across the stages of
galaxy evolution. The column density distribution function, $f(\nhi)$
is one way to characterise this distribution. $f(\nhi)$ is defined as
the probability of intersecting a particular column density along a
random line-of-sight per unit distance. It describes the
cross-sectional area of sky occupied by a particular column density at
a given redshift. Traditionally $f(\nhi)$ is measured by \Lya\
absorption in the continuum spectra of background QSOs. At high
redshift $f(\nhi)$ is well measured over 10 orders of magnitude in
column desnity (eg. Tyler 1987; Petitjean et al. 1993). At low
redshift, since the \Lya\ transition is in the UV and the occurrence
of high column density absorbers is low, we have measured the neutral
hydrogen column density distribution using 21-cm \HI\ emission. The
purpose of this investigation is to calculate $f(\nhi)$ using a range
of galaxy \HI\ masses and determine their contribution to the column
density distribution function.

In the last few years it has become apparent that Damped Lyman
$\alpha$ (D\Lya) systems are not necessarily caused by the \HI\ disks
of large spiral galaxies (eg. Rao \& Turnshek 1998; Bowen, Tripp, \&
Jenkins 2001). Regardless of this, the paradigm of large spirals being
responsible for the majority of the cross-section of column densities
\nhi\ $>10^{19}$ at z=0 remains.

\section{Motivation and Justification}

According to Dickey \& Lockman (1990) ``The agreement between
ultraviolet and radio estimates of \nhi\ is excellent and somewhat
astonishing, considering that the angular resolution, experimental
techniques, and telescopes differ in all respects''. This statement
refers to a comparsion of high latitude Galatic \HI\ 21-cm
measurements compared with \Lya\ absorption towards distant stars (in
the same direction) in the range $10^{20} < \nhi\ < 10^{21}$ \cm. One
concern remains that high column density gas may clump on scales less
than the beam size. This has been addressed by Rao \& Briggs (1993) in
their calculation of the contribution of high \nhi\ gas to the \HI\
mass function (which is not affected by beam size) compared with the
integral of mass from the $f(\nhi)$ cross-section. They find the
integral of mass in $f(\nhi)$ is equivalent to that from the mass
function at high \nhi. They concluded that any high \nhi\ column
density gas missed due to a large beam size is negligible.

The advantages of using 21-cm observations to measure column density,
rather than \Lya\ absorption are:
\begin{itemize}
\item For z$\lesssim$1.65, \Lya\ requires space based observations. 
\item The space density of \Lya\ absorbers at low redshift is
small. Indeed, Rao \& Briggs (1993) estimated that 1000 random sight lines
would be required to provide 10$\pm$3 D$\Lya$ systems. According to
Rao and Turnshek (2000) 23 D$\Lya$ systems are known at z$<$1.65. 
\item Significant evolution of the column density distribution is still
expected between z$\sim$1 and z=0, so it is important provide an
anchor point at z=0.
\item Foreground galaxies may hinder QSO identification. Extinction
and reddening by dust in galaxies means that QSOs may be preferentially
found away from galaxies (0striker \& Heisler 1986).

\end{itemize}

\section{Comparison to previous work}

Previous calculations of the column density distribution function from
21-cm emission has published by Rao \& Briggs (1993) and Zwaan,
Verheijen, \& Briggs (1999). The study by Rao \& Briggs used Arecibo
observations of 27 large spiral galaxies. Zwaan et al. on the other
hand used synthesis observations of Ursa Major cluster members.  The
advantage of the method presented here is that a range of \HI\ mass
galaxies are including in the sample from the whole sky, rather than a
galaxy cluster, which are known to \HI\ deficient (eg. Solanes et
al. 2001). The benefit of using a galaxy cluster is that the spatial
resolution is the same in each map. We have preserved this feature by
selecting galaxies within a tight velocity range. We have also used an
\HI\ mass function for normalisation which is more appropriate than an
optical luminosity function, used by the two previous studies.

\section{Observations}

Galaxies for this study were selected from the \HI\ Parkes all sky
survey (HIPASS) Bright Galaxy Catalogue (Koribalski, 2000). HIPASS is
a blind survey for \HI\ covering the entire southern sky, thus
providing a complete \HI\ selected sample. The 3$\sigma$ \HI\ mass
limit of HIPASS is $10^{6} D^{2}_{Mpc}\msun$ and the 3$\sigma$ \HI\
column density limit is $4\times10^{18}$ \cm. To provide a
representative range of \HI\ mass, 36 galaxies were chosen randomly to
fill six half decade \HI\ mass bins from $10^{7.5}$ to $10^{10.5}$ \msun.
Galaxies were restricted to those with helocentric velocities in the
range 500 to 1700 \kms\ so that the spatial resolution across the
sample remained comparable across the sample. The gridded FWHP
beamwidth of HIPASS is 15.5\arcmin\ which means that most galaxies are
poorly resolved. Since the HIPASS maps provide a column density spread
over the entire beam area, higher resolution mapping was required to
measure distribution of column desnity. Each galaxy was observed with
the Australia Telescope Compact array (ATCA) with a minimum FWHM beam
size of 30\arcsec. The typical 3$\sigma$ \HI\ column density limit for
the observations was $\sim 3\times10^{18}$ \cm.

\section{Data Analysis}
The brightness temperature measured for each pixel in the synthesis
map, T$_{b}$ is converted to column density by the relationship:
\begin{equation}
\nhi = 1.823\times 10^{18}\int T_{b}dv \hspace{3mm} \text{cm}^{-2} .
\end{equation}
This calculation assumes that the gas is optically thin. Dickey \&
Lockman (1990) used typical Galactic \HI\ parameters (T${_s}\sim50$K
and $\Delta v\sim10$\kms) to show that the optical depth is greater
than 1 for $\nhi > 10^{21}$ \cm. Consequently the column density may be
underestimated as \nhi\ approaches $10^{21}$ \cm. The column density
distribution function is then calculated by
\begin{equation}
f(\nhi)=\frac{c}{H_0}\frac{\phi\Sigma(\nhi)}{d\nhi},
\end{equation}
where $\Sigma(\nhi)$ is the area in Mpc$^2$ occupied by a particular
column density, in the range \nhi\ to $\nhi+d\nhi$. The function is
normalised by the space density $\phi$ taken from a HIPASS \HI\ mass
function (Kilborn, 2000). The \HI\ mass function has a faint end slope
of $\alpha = 1.52$ and an characteristic \HI\ mass of
$M_*=1.14\times10^{10}\msun$. An $H_o$ of 75 \kms\ Mpc$^{-1}$ is
adopted to match that of the \HI\ mass function and a value of
$d\nhi=0.2$ dex is used throughout. The distance to each galaxy was
calculated assuming the velocity represents the Hubble flow and
corrected for Local Group velocity. Uncertainty in the distance is a
large source of error; it affects both \mhi\ and the area function,
$\Sigma(\nhi)$. $f(\nhi)$ is plotted in Figure 1 with a power law
fitted with the form $f(\nhi) = B\nhi^{-\beta}$. From the figure it is
evident that low \HI\ mass galaxies dominate $f(\nhi)$ at column
densities in the range $3\times 10^{18}$ to $2\times 10^{20}$ \cm. At
the D\Lya\ cutoff of $2\times 10^{20}$ \cm\ both mass types contribute
equally. For $\nhi\ > 2\times 10^{20}$ \cm, M$_*$ galaxies dominate
the column density distribution function.
 
\section{Discussion}
From the \HI\ mass function we know that galaxies with $10^{6} < \mhi\
< 10^{8}$ \msun\ contribute only 6\% of the total \HI\ mass density at
z=0. Indeed most of the \HI\ mass in the local Universe is contained
in galaxies with $10^{9} < \mhi\ < 10^{10.25}$ \msun\ (Kilborn,
2000).  Even though their contribution to the overall \HI\ mass
density is minimal, the cross-sectional area presented by low \HI\
mass galaxies for moderate column densities is significant. This
result is interesting as it is not intuitively obvious. The space
density of low \HI\ mass galaxies is sufficient to push their
cross-sectional area at moderate \nhi\ above that of M$_{*}$
galaxies. The morphology of the galaxies in the lowest mass bin are
generally Irregular type. In contrast to those in the M${_*}$ mass bin
which are generally Early type spirals.

This result has implications for the interpretation of the column
density distribution function at all redshifts. If systems with \mhi\
= $10^{7.5}$ \msun\ are responsible for a significant fraction of the
moderate column density cross-section at low redshift, they may also
cause a significant fraction of absorption systems at high
redshift. As noted above, the typical morphology of these systems is
Irregular. Normal spiral galaxies are known to decrease their star
formation rate (SFR) with time, yet recent evidence suggests that
Irregular galaxies experience an increasing SFR with time (Xu et
al. 1994).  Therefore galaxies associated with moderate \nhi\ \Lya\
absorption systems at high redshift may have been missed due to low
optical luminosity.

\section{Conclusion}
We have shown that at z=0, the contribution of low \HI\ mass galaxies to the
column density distribution function is greater than M$_*$ galaxies
for \nhi\ up to the D\Lya\ cutoff. This result implies that
moderate \nhi\ \Lya\ absorption systems may be caused by a range of
galaxy types and not just large spiral galaxies as originally
thought.

\begin{figure}
\plotone{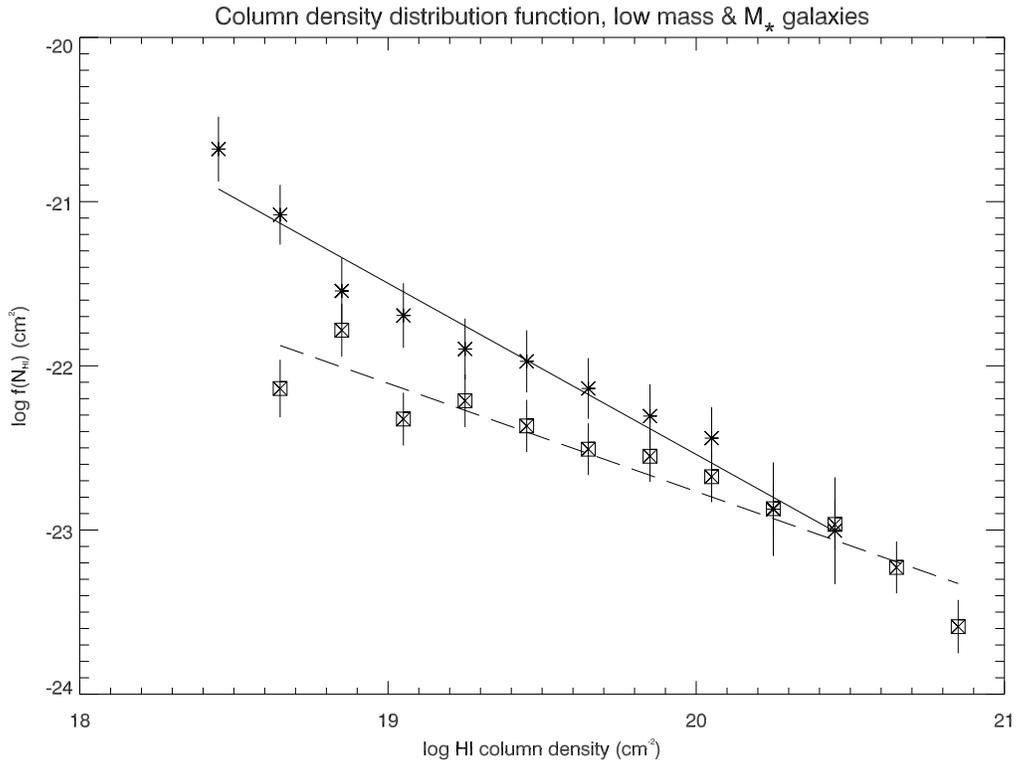}
\label{fig:cddf}
\caption{ Column density distribution function at z=0
for two different \HI\ mass galaxy types. $f(\nhi)$ from low \HI\ mass
galaxies ($10^{7.5} < \mhi\ < 10^8$ \msun) are represented by the
solid line and asterisk data points. $f(\nhi)$ from M$_{*}$ galaxies
are represented by the dashed line and square data points.}
\end{figure}

\end{document}